# Solution-Processed Inks with Fillers of NbS₃ Quasi-One-Dimensional Charge-Density-Wave Material

Tekwam Geremew[1,2], Maedeh Taheri[1,2], Nicholas Sesing[3], Subhajit Ghosh[1,2],
Fariborz Kargar[4] *, Tina T. Salguero[3], and Alexander A. Balandin[1,2]*

[1]Department of Materials Science and Engineering, University of California, Los Angeles, California 90095 USA

[2]California NanoSystems Institute, University of California, Los Angeles, California 90095 USA

[3]Department of Chemistry, University of Georgia, Athens, Georgia 30602 USA

[4]Materials Research and Education Center, Department of Mechanical Engineering, Auburn University, Auburn, Alabama 36849 USA

---

* Corresponding authors: fkargar@auburn.edu (F.K.); balandin@seas.ucla.edu (A.A.B.); web-site: https://balandin-group.ucla.edu/

**Abstract**

We report on the solution processing and testing of electronic ink comprised of quasi-one-dimensional $NbS_3$ charge-density-wave fillers. The ink was prepared by liquid-phase exfoliation of $NbS_3$ crystals into high-aspect ratio quasi-1D fillers dispersed in a mixture of isopropyl alcohol and ethylene glycol solution. The results of the electrical measurements of two-terminal electronic test structures printed on silicon substrates reveal resistance anomalies in the temperature range of ~330 K to 370 K. It was found that the changes in the temperature-dependent resistive characteristics of the test structures originate from the charge-density-wave phase transition of individual $NbS_3$ fillers. The latter confirms that the exfoliated $NbS_3$ fillers preserve their intrinsic charge-density-wave quantum condensate states and can undergo phase transitions above room temperature even after chemical exfoliation processes and printing. These results are important for developing "quantum inks" with charge-density-wave fillers for the increased functionality of future solution-processed electronics.

**Keywords:** one-dimensional materials; charge density waves; printed electronics; solution-processed electronics; quantum materials, $NbS_3$

# 1. INTRODUCTION

Printed electronics with inks of solution-processed low-dimensional materials, such as graphene and transition metal dichalcogenides (TMDs), have attracted significant interest owing to the potential for offering versatile device applications at lower manufacturing costs[1–4]. Printing adopts an additive manufacturing technique to deposit micro and nanoscale materials on substrates with different rigidity through various printing technologies [5–8]. Among the printing approaches, inkjet printing is considered one of the promising technologies owing to its mask-free printing requirements, large scale, and high manufacturing throughput [9–13]. Generally, in inkjet printing, solution-based inks with fillers of electrically conductive, insulative, or semiconductive nature are deposited onto flexible polymer or conventional silicon substrates [14–17]. For example, graphene-based conductive inks, printed on flexible substrates, have given rise to device applications such as humidity sensors, electrochemical sensors, transistors, photovoltaic cells, and wireless data communications[18–21]. Other two-dimensional (2D) TMDs, such as $MoS_2$ and $WS_2$ ink devices, have been widely employed for printed optoelectronics devices, sensor applications, logic gates, and energy storage devices [22–24].

Although printing with inks of 2D and other low-dimensional semiconductors offers exciting opportunities for novel electronics, it suffers from a fundamental shortcoming. That is the overall electric transport in any printed channel with semiconducting fillers will be dominated by the electron hopping mechanism [25–27]. In a printed device with inks of semiconducting fillers, the electric conductivity in individual fillers is governed by band-type charge transport. However, at the filler-filler interfaces, charge hopping will become the dominant transport mechanism under any conditions. This situation explains why printed devices with semiconducting fillers demonstrate much lower mobilities compared to devices fabricated with conventional cleanroom techniques [28,29]. Furthermore, other undesired effects such as defects and impurities are usually induced in the fillers during the ink preparation and make the situation even worse. As a result, it is challenging to maintain the original properties of the pure crystalline semiconductor materials in their disordered form within the ink. To take advantage of opportunities offered by printing and to overcome the challenges described above, one can think of utilizing functional low-dimensional materials with unique properties, such as charge-density-wave (CDW) materials.

The CDW phenomena are the quantum condensate phase effects, observed in some quasi-one-dimensional (1D) and quasi-2D materials [30–33]. It manifests as a periodic modulation of electron charge density, coupled with a corresponding periodic distortion of the underlying crystal lattice [33]. This effect results in anomalies in temperature-dependent charge transport characteristics that can be harnessed for electronic devices, such as voltage-controlled oscillators [34,35]. To design CDW-based printed devices, the first challenge to be addressed is whether such materials can preserve their CDW quantum properties after the chemical processes required for ink preparation.

Here, we report on the preparation of functional inks with one-dimensional (1D) $NbS_3$ CDW material as filler. $NbS_3$ belongs to the family of transition metal trichalcogenides (TMTs), a class of 1D van der Waals (vdW) compounds where M is a transitional metal and X is a chalcogen atom [36]. Among the $MX_3$ materials, $NbS_3$ has attracted much attention owing to its polymorphism and the presence of the CDW quantum condensate phases above room temperature (RT) [37–39]. Similar to other $MX_3$ materials, $NbS_3$ has a chain-like structure with strong covalent bonds in the chain direction. Chains are arranged as bilayers through Nb–S interactions, and these layers are held together by relatively weak vdW forces that allow top-down nanostructuring *via* mechanical or liquid-phase exfoliation (LPE) techniques [37,40]. This is a key feature because the resulting quasi-1D $NbS_3$ nanowires have high aspect ratios and mechanical flexibility, properties that allow electrical percolation to be reached at lower concentrations [41,42]. This characteristic is crucial for the preparation of inks with desired thermophysical properties and for achieving reasonable electrical conductivity after the printing of a few layers.

## 2. MATERIAL CHARACTERIZATIONS

The various $NbS_3$ polymorphs differ in Nb–Nb bonding along the chains as well as in the specific packing arrangements of chains and bilayers. Figure 1 (a) shows the basic chain and bilayer structures of $NbS_3$-II and $NbS_3$-V, both of which are metallic.[39,43] The polymorphs belong to $P2_1/m$ space group with monoclinic crystal structures with a uniform $Nb^{4+}$ chain distance of ~3.35 Å [39]. For $NbS_3$-II, its complex unit cell structure contains four types of symmetry-related trigonal prismatic (TP) chains, resembling the structure of $NbSe_3$ [44]. Polymorph II, in particular, is known to exhibit three CDW phases at $T_{P0}$ ~ 460 K, $T_{P1}$ ~330–370 K, and $T_{P2}$ ~150 K with considerable change in its resistive profile [44–46]. For this study, we used NbS$_3$ crystals synthesized by the chemical vapor transport (CVT) technique; growth from the elements employed a 973 K – 943 K heating gradient and excess sulfur for transport [39,47]. Figure 1 (b) presents the scanning electron microscopy (SEM) imaging and the corresponding energy dispersive x-ray spectroscopy (EDS) mapping of the as-grown wire-like NbS$_3$ crystals.

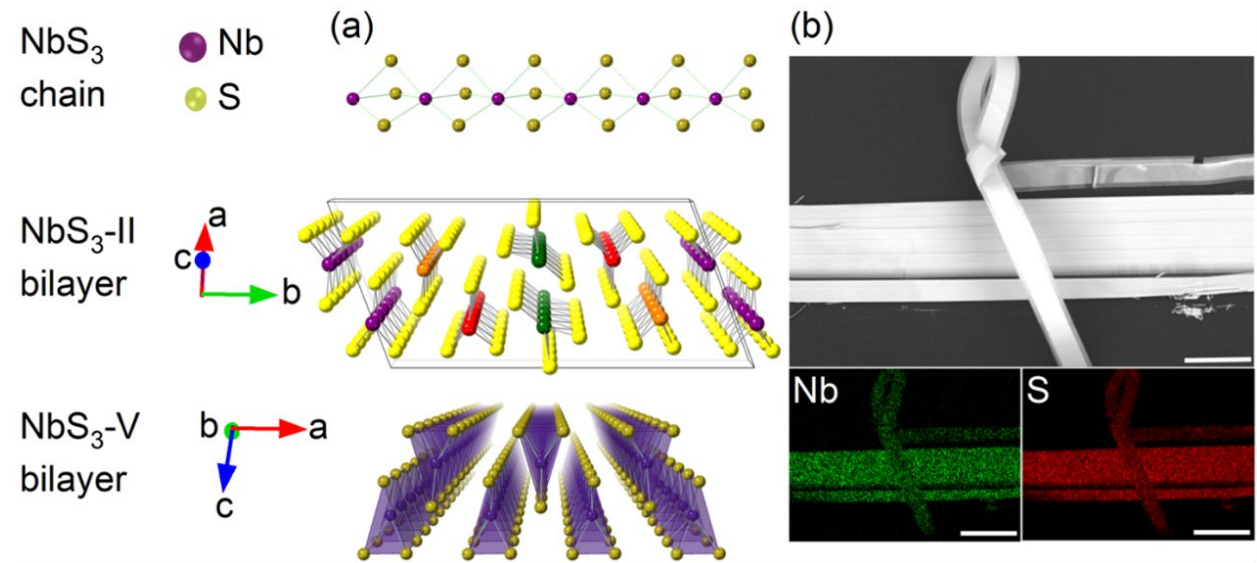

[**Figure 1**: (a) The crystal structures of the NbS$_3$ single chain and bilayer structures of NbS$_3$-II and NbS$_3$-V polymorphs. (b) The representative SEM imaging and EDS elemental mapping of the as-grown NbS$_3$ crystals. The scale bar is 50 μm.]

To verify the electrical characteristics of these NbS$_3$ crystals, we first measured the temperature-dependent electrical resistance of devices fabricated using *clean* mechanically exfoliated NbS$_3$ nanowires. NbS$_3$ whiskers were mechanically exfoliated into high aspect-ratio thin nanowire chains on top of a clean SiO$_2$/Si substrate using Nitto tape. To protect the NbS$_3$ nanowire from environmental exposure, a thin *h*-BN layer was transferred on top of the nanowire using an all-dry transfer method. Contacts were made by patterning the device using e-beam lithography (EBL), followed by etching the *h*-BN protective layer using reactive ion etching (RIE) and finally, depositing Ti/Au metals with 10 nm/90 nm thickness in an e-beam evaporation (EBE) chamber. Figure 2(a) shows an optical microscopy image of a representative *h*-BN-capped NbS$_3$ nanowire device on a $SiO_2$/Si substrate. The electrical properties of the $NbS_3$ device channel were measured in a two-terminal configuration at different temperatures.

The measurements were conducted inside a cryogenic probe station (Lakeshore TTPX) using a semiconductor parameter analyzer (Agilent B1500A). Figure 2 (b) shows the current-voltage (*I-V*) behavior at room temperature. As seen, the *I-V* plot follows a linear ohmic behavior at lower biases with super-linear characteristics after a particular bias point, referred to as the threshold voltage, $V_{th}$. In the context of CDW phenomena in quasi-1D materials, the linear part of the electric transport behavior is attributed to the current carried by individual electrons. The nonlinearity in *I-V* at and beyond $V_{th}$ is attributed to the collective current produced by depinning and sliding of CDW condensate in the incommensurate CDW phase of the material. The total current, $I_{total}$, is a summation of linear current, $I_L$, and collective CDW current $I_{CDW}$, *i.e.*, $I_{total} = I_L + I_{CDW}$.[48] The temperature-dependent *I-V* plots of the device, shown in Figure 2 (c), in the temperature range between 150 K and 400 K, also followed similar threshold field-dependent *I-V* behavior, indicating the presence of incommensurate CDW phase across the measured temperature range. More details regarding the depinning of CDW phases in different quasi-1D materials have been reported by some of us and others elsewhere and it is beyond the scope of this study [47,49,50].

The resistance, *R*, of the $NbS_3$ nanowire device at different temperatures was extracted from the linear ohmic part of the *I-V* data. The results are presented in Figure 2 (d) (blue symbols) in the semi-log scale as a function of the inverse temperature, *1000/T*. As seen, the resistance increases with the inverse temperature. Two distinguishable sharp changes can be detected in the range of ~320 K to 370 K, around the CDW phase transition temperature, $T_{P1}$. This region is shaded with green color in the Figure for a better visualization. The resistance changes around $T_{P1}$ can be attributed to the CDW Peierl's transition [44]. The derivate of the resistance as a function of temperature, shown by red symbols in Figure 2 (d), points out more clearly a prominent peak at 370 K accompanied by a smaller shoulder around ~320 K. This behavior for $NbS_3$ is similar to the transition temperature reported previously for polymorph II and is consistent with a transition to the incommensurate CDW phase in our device [51,52].

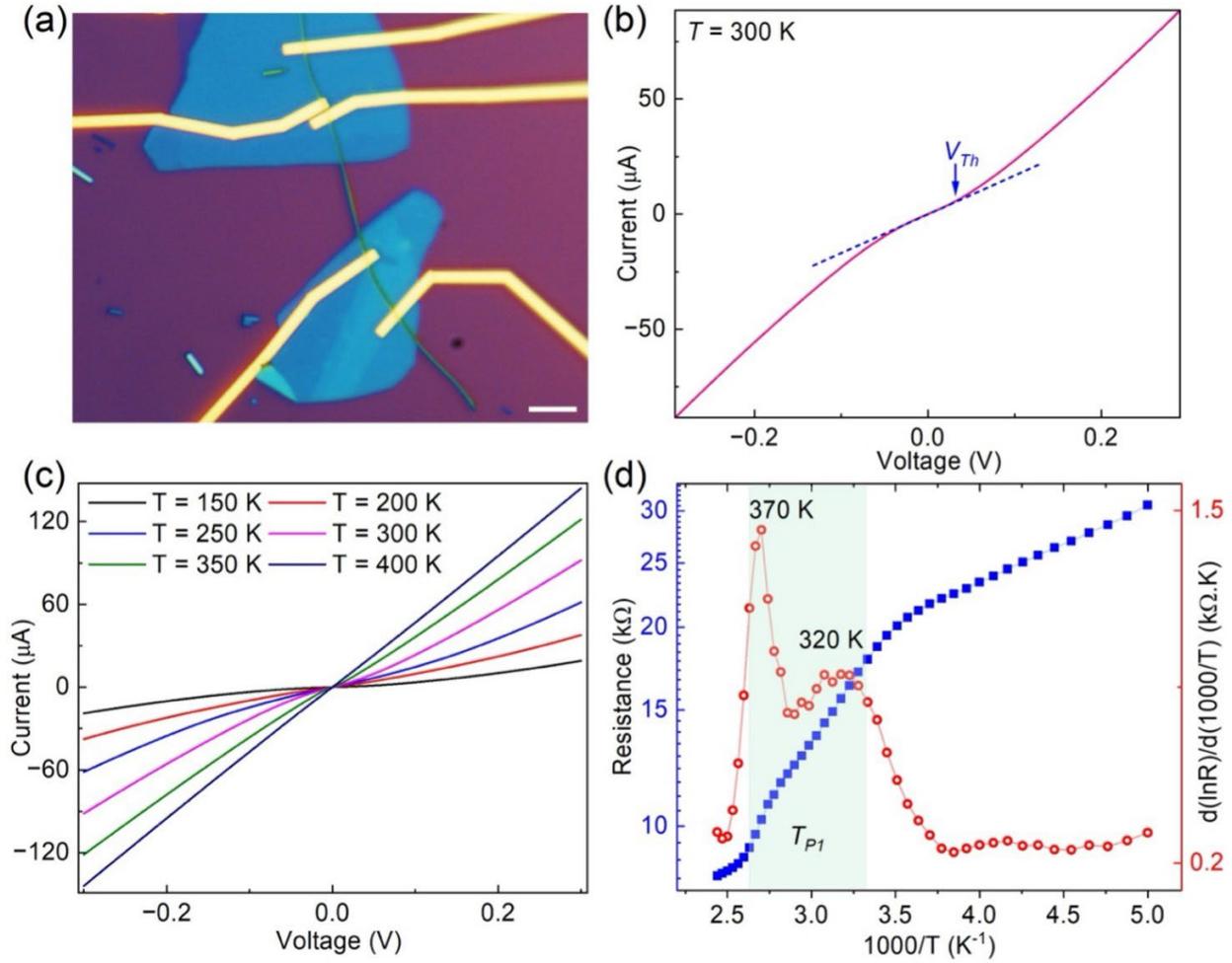

[**Figure 2**: (a) Optical microscopy image of a representative NbS$_3$ device. The nanowires were isolated using mechanical exfoliation and capped with an *h*-BN layer to protect the channel from environmental exposure. The scale bar is 5 μm. (b) The *I-V* characterization of the NbS$_3$ device at room temperature. The start of nonlinearity, which has been shown with an arrow is attributed to the depinning and sliding of the CDW quantum condensate and its contribution to total conduction. (c) The temperature-dependent *I-V* characterization of the NbS$_3$ device in the temperature range of 150 K to 400 K. (d) The electric resistive characteristic of the device plotted as a function of inverse temperature in a semi-logarithmic scale, in the temperature range of 150 K to 400 K. The blue and red symbols show the resistance and its temperature derivative, respectively. The derivative data shows a sharp peak at 320 K accompanied by a shoulder at 370 K in the vicinity of the CDW phase transition, $T_{P1}$.]

## 3. INK PREPARATION AND PRINTED DEVICE FABRICATION

We now focus on the central aspect of this study which is the formulation procedure for ink containing $NbS_3$ fillers, tailored for electronic printing applications. We employed an ultrasonication-based liquid-phase exfoliation (LPT) technique to prepare the ink with exfoliated quasi 1D $NbS_3$ fillers. In this technique, a certain amount of the bulk material is added to the proper solvent, and the solution undergoes ultrasonication for a certain period of time. Exfoliation of the bulk material occurs due to micro-jets and shock waves generated during the collapse of microbubbles induced by liquid cavitation acting on the layered materials [53,54]. Liquid-phase exfoliation is regarded as one of the most cost-effective techniques for the large-scale exfoliation of vdW materials owing to its straightforward mechanics and high adaptability [55–57].

To break down bulk $NbS_3$ crystals into thinner nanowires, various solvents were tested, which included acetone, isopropyl alcohol (IPA), and dimethylformamide (DMF). DMF exhibits the highest efficiency among these solvents, effectively exfoliating more crystals in a relatively short period. Figures 3 (a, b) show the optical microscopy and scanning electron microscopy (SEM) images of the CVT-grown bulk $NbS_3$, respectively. Note the needle-like structure of the material, even in the bulk form. To prepare the ink, 5 mg of the bulk $NbS_3$ is added into a 20 ml high-boiling-point DMF solvent (Figure 3 (c)). The mixture underwent exfoliation for 6 hours in a low-power bath sonicator (5510 Branson). As seen in Figure 3 (d), after sonication, the color of the mixture changed from clear to dark brown, confirming an efficient exfoliation and dispersion of bulk $NbS_3$ in DMF. To separate the supernatant and precipitate components, the solution was transferred to a centrifuge tube and underwent centrifugation for 1 hour at 2000 rpm. Figure 3 (e) shows the mixture after the centrifugation process. The precipitate component is collected to be used in the final step of the ink preparation which will be discussed next.

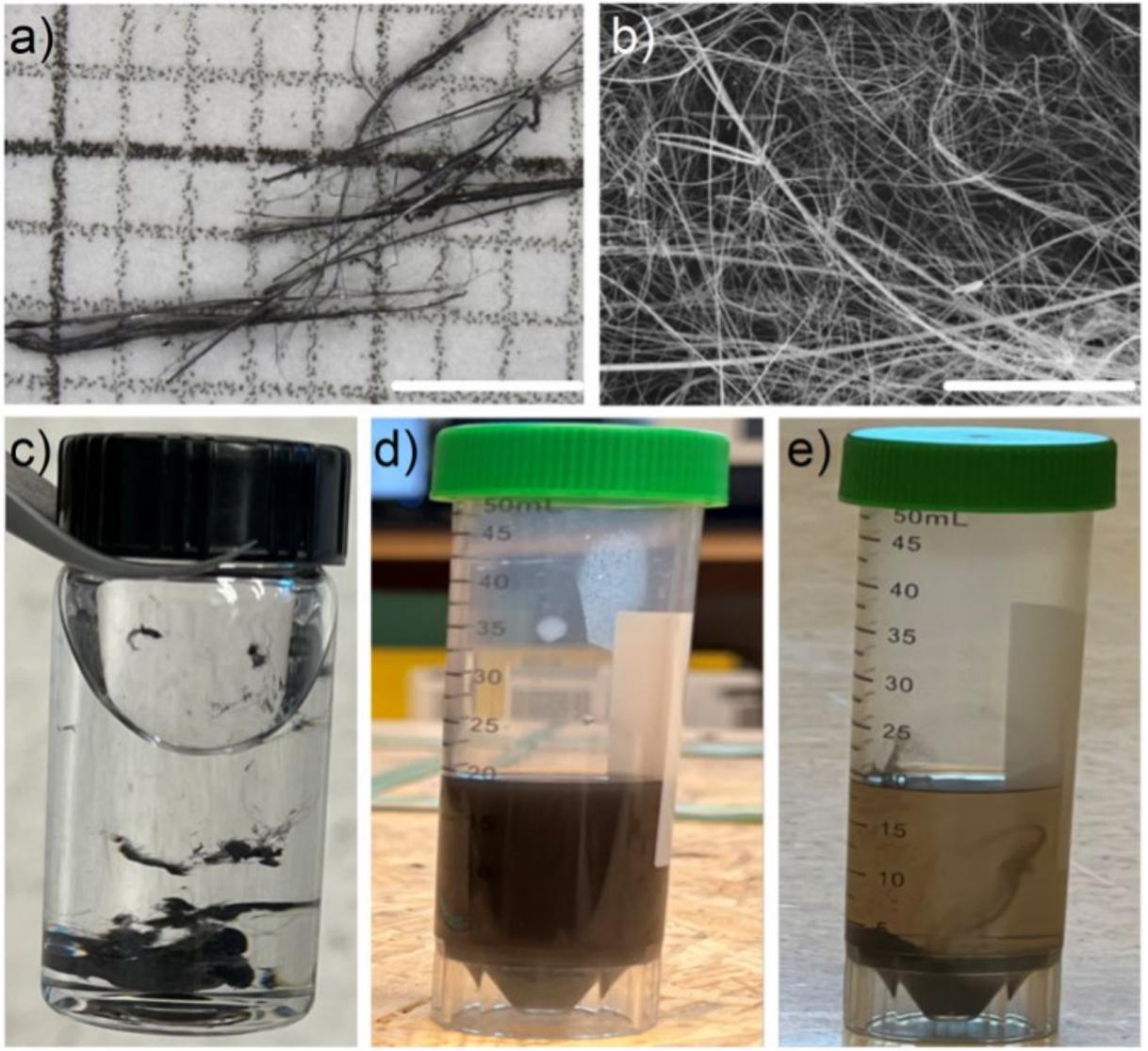

[**Figure 3**: (a) Optical microscopy, and (b) SEM images of the bulk CVT-grown NbS$_3$ material. Note the quasi-1D needle-like structures of NbS$_3$ even in the bulk form. The scale bar in the optical and SEM images are 2.5 mm and 50 μm, respectively. (c) Mixture of NbS$_3$ and DMF as solvent before sonication. (d) The solution of NbS$_3$ in DMF after undergoing bath sonication. Note the change in the color of DMF solvent from colorless to dark brown confirming an efficient exfoliation of NbS$_3$. (e) The same mixture after the centrifugation process. The precipitate part of the mixture, settled at the bottom of the vial, is used for final ink preparation.]

In printing electronics, the thermophysical properties of the ink and the size of the injecting nozzle are key factors in the formation of proper ink droplets with desirable characteristics.[58–62] The droplet formation criteria is determined by the $Z$-number, $Z = \sqrt{\zeta \rho a}/\mu$, a dimensionless parameter in which $\zeta$, $\rho$, and $\mu$ are respectively, the surface tension, density, and the dynamic viscosity of the ink solution, and $a$ is the diameter of the injecting nozzle.[63,64] A $Z$-number range of $1 \leq Z \leq 14$ generally results in good droplet formation and its uniform distribution at the intended regions on the substrate.[65] Inks with $Z < 1$ or $Z > 14$ result in alongated ligamaents or satellite droplet formation which would result in undesirable printing patterns.[63,66–69] To tune the properties of our ink, we collected the precipitate part of the liquid-phase exfoliated NbS$_3$ and transferred it to a secondary solvent of 5 ml isopropyl alcohol (IPA). The solution underwent a 10-minute bath sonication for uniform filler dispersion and to prevent filler agglomeration. Despite the advantageous properties of IPA such as low boiling point and high evaporation rate, IPA-based solutions tends to have high surface tension and low viscosity, resulting in the creation of undesirable "coffee-rings" after the drying of the ink on the substrate.[15] To resolve these issues, an equal volume of ethylene glycol (EG) was added to the IPA solution containing NbS$_3$ fillers. EG has a higher boiling and its addition to the IPA-based solution increases the solvent's boiling point.[66,67] Mixing of two solvents with different boiling points and surface tensions prevents the formation of coffee-rings which can be explained by the Marangoni effect.[70,71] Figure 4 (a) shows the resultant ink for printing.

Figure 4 (b) illustrates a schematic representation of the two-terminal printed device used in this study. Prior to initiating the printing, electrodes were made by applying silver (Ag) paste on a Si/SiO$_2$ substrate with a printed channel length of ~1.6 mm. The electrodes were annealed for 40 min at 70 °C using a hot plate. A piezo-driven inkjet printer (Hyrel 30M 3D) with a nozzle diameter of $a = 0.21$ mm was used to print the ink solution containing NbS$_3$ fillers as the conductive channel between the electrodes. The bed temperature of the printer was kept at 70 °C to expedite the solvent evaporation and ink-drying process and also to mitigate the coffee-ring effect by enhancing the movement of fillers within the ink droplet after deposition. Figure 4 (c) presents a long-shot SEM image of the ink channel. As seen, no coffee-rings were observed in the channel confirming the high quality of the printed device. Figure 4 (d) displays a magnified image of the channel. A network of quasi-1D NbS$_3$ fillers with different aspect ratios, deposited uniformly in the channel, can

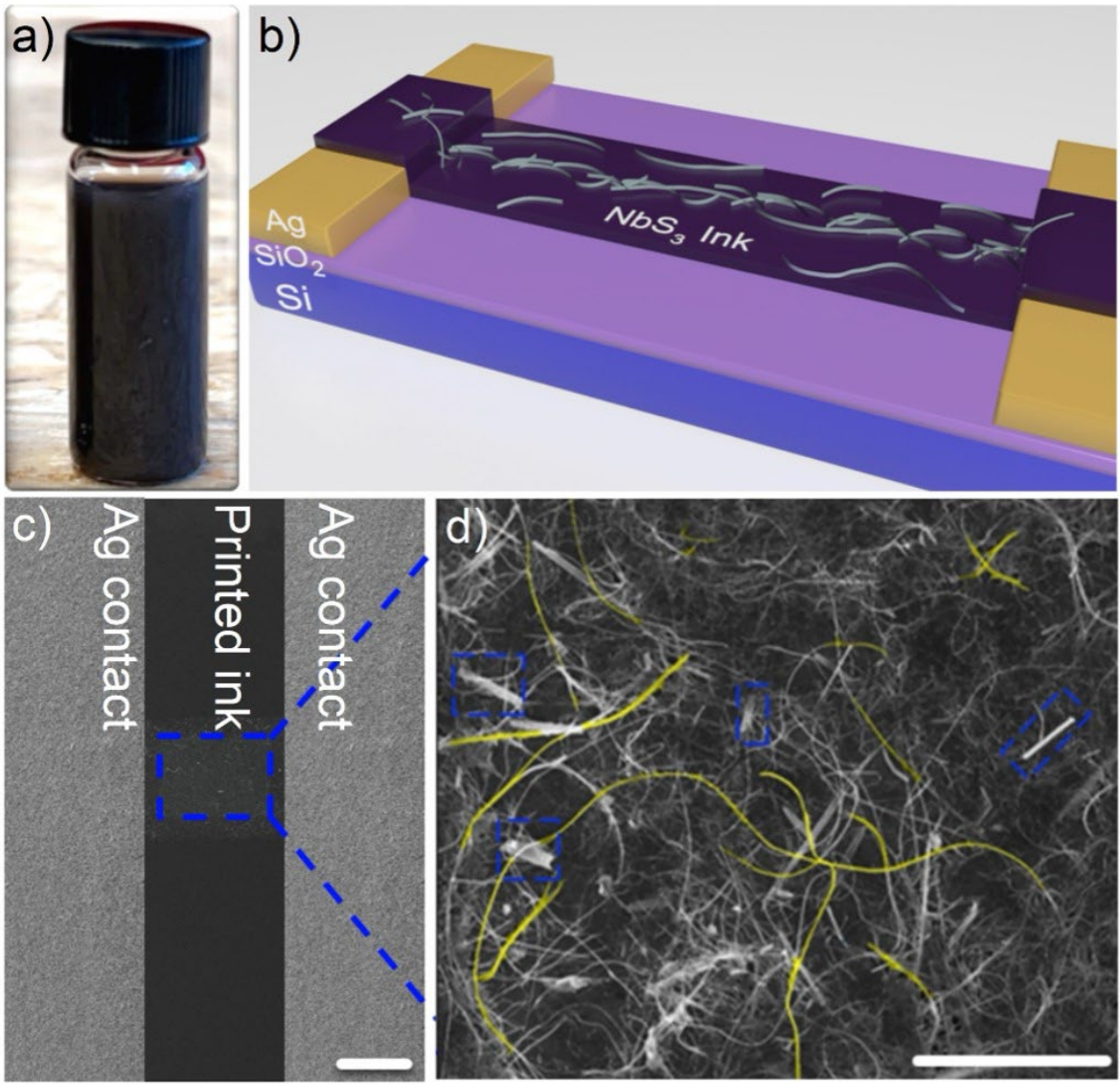

[**Figure 4**: (a) NbS₃ ink with ethylene glycol and IPA as solvents with mixing volume ratios of (1:1). (b) Schematic showing the basic structure of the printed NbS₃ device. (c) SEM image of the printed device showing the NbS₃ ink channel and silver paste (Ag) as electrodes. (d) High-resolution zoomed-in SEM image of the NbS₃ printed ink. The close-up view of the printed NbS₃ material exhibits finer details of the ink structure. The "quantum ink" consists of quasi-1D NbS₃ fillers of different polymorphs of "rigid needle-like" NbS₃-I or IV and "flexible hair-like" NbS₃-II, with the latter being the predominant phase. Likely NbS₃-I and NbS₃-II fillers are identified in blue and yellow colors. The dashed blue box in (c) is not to scale with the magnification of image shown in (d). The scale bars in (c) and (d) are 20 μm and 200 nm, respectively.]

be seen in this image. Note that some fillers appear as rigid nanoribbons, enclosed by blue dash lines, while the majority of the fillers have flexible quasi-1D nanowire structures (shown in yellow color). The rigid, needle-like wires most likely are of type-I or type-IV polymorphs, whereas the fine, hair-type wires more closely resemble type-II.[38,47] Given that the ink channel consists of mostly flexible type-II $NbS_3$ fillers, one should expect to see signatures of CDW transition in the printed device similar to the data presented in Figure 2 (d).

## 4. RESULTS AND DISCUSSION

To examine whether the $NbS_3$-based ink has preserved the CDW properties, we measured the electrical resistance of the printed device in the temperature range of 295 K to 393 K in the heating cycle. We employed the standard two-probe electrical resistance measurements using a multimeter (Fluke 289). The device was heated using a hot plate (Corning PC-4000). A thermocouple was attached to the substrate close to the printed channel to read the temperature of the device at each measurement. Figure 5 (a) presents the resistance of the $NbS_3$ ink channel device, $R$, as a function of inverse temperature, *1000/T*, in a semi-log scale. The channel resistance was measured to be ~2 MΩ at RT. As depicted in Figure 5 (a), the resistance of the channel increases with temperature rise with two noticeable variations in the slope occurring at T~ 325 and T ~ 375. These temperatures correspond to the maxima of the *dR/dT* data shown in Figure 2 (d) and agree well with the CDW transition temperatures reported for single-crystal $NbS_3$-II. These results suggest that $NbS_3$-II preserves its CDW properties despite undergoing liquid-phase exfoliation and printing processes. One should note that defects and disorder are detrimental to any coherent quantum states. The latter explains the significance of preserving the CDW phases in the prepared "quantum ink".

To verify that the features observed in the electrical resistance of the prepared inks are associated with the intrinsic properties of $NbS_3$, we prepared a reference graphene-based ink following the same recipe we used to prepare the $NbS_3$-based ink. A similar two-terminal device with the same channel length and contacts was printed and its temperature-dependent electrical resistance was measured. The results of the temperature-dependent resistance for the $NbS_3$ and graphene inks are presented in Figure 5 (b). As seen, the electrical resistance of the graphene ink gradually decreases with heating, showing no detectable anomalies in the slope of its resistance at any examined temperatures. In contrast, the NbS$_3$-II ink exhibits an obvious change in resistive characteristics at T ~ 325 K and T~375 K. Based on the comparison with the properties of individual NbS3 fillers (see Figure 2 (d)) and with the properties of the reference ink (Figure 5 (b)), it can be inferred that the non-trivial resistance anomalies observed at these temperatures stem from the CDW phase transitions of NbS$_3$-II fillers present within the ink. It is interesting to note that although the NbS$_3$-ink used in this study consists of fillers of different polymorphs, the CDW characteristics are still observable in the printed device. This finding offers an exciting opportunity for the development of technologically viable and economically feasible CDW-based printed devices.

The temperature-dependent resistive behavior of printed devices can be explained as an interplay between the CDW phase transition of the intrinsic material and a complex transport mechanism of the disordered NbS$_3$ ink channel. Prior studies and the resistive behavior of the device with single-crystal NbS$_3$-II as an active channel, presented in Figure 2 (d), have shown that NbS$_3$ crystals undergo CDW phase transitions within a wide temperature span, encompassing the range of 325 K to 370 K. In the ink device, a sharp change in resistance is also observed in a similar temperature interval, between $T$ = 314 K and 376 K. However, unlike the device with single-crystal NbS$_3$-II, the resistance change is not continuous but slows down in intermediate temperatures. The overall carrier conduction process in disordered systems such as inks is governed by hopping charge transport between the delocalized sites of the adjacent fillers.[72] However, in inks with CDW fillers, the transport is also affected by CDW transitions of individual NbS$_3$ fillers in the percolated network of fillers. Other factors, such as random interconnections and interfaces between NbS$_3$ fillers, defects, and impurities introduced during the LPE process and the subsequent printing steps also influence the transport properties [73,74]. As a result, charge carrier transport through random networks can be significantly different from that of the transport in single-crystal materials. This explains the slower resistance change in the intermediate temperatures, which deviates from the continued sharp resistive change observed in single-crystal NbS$_3$-II in the temperature across the CDW phase transition.

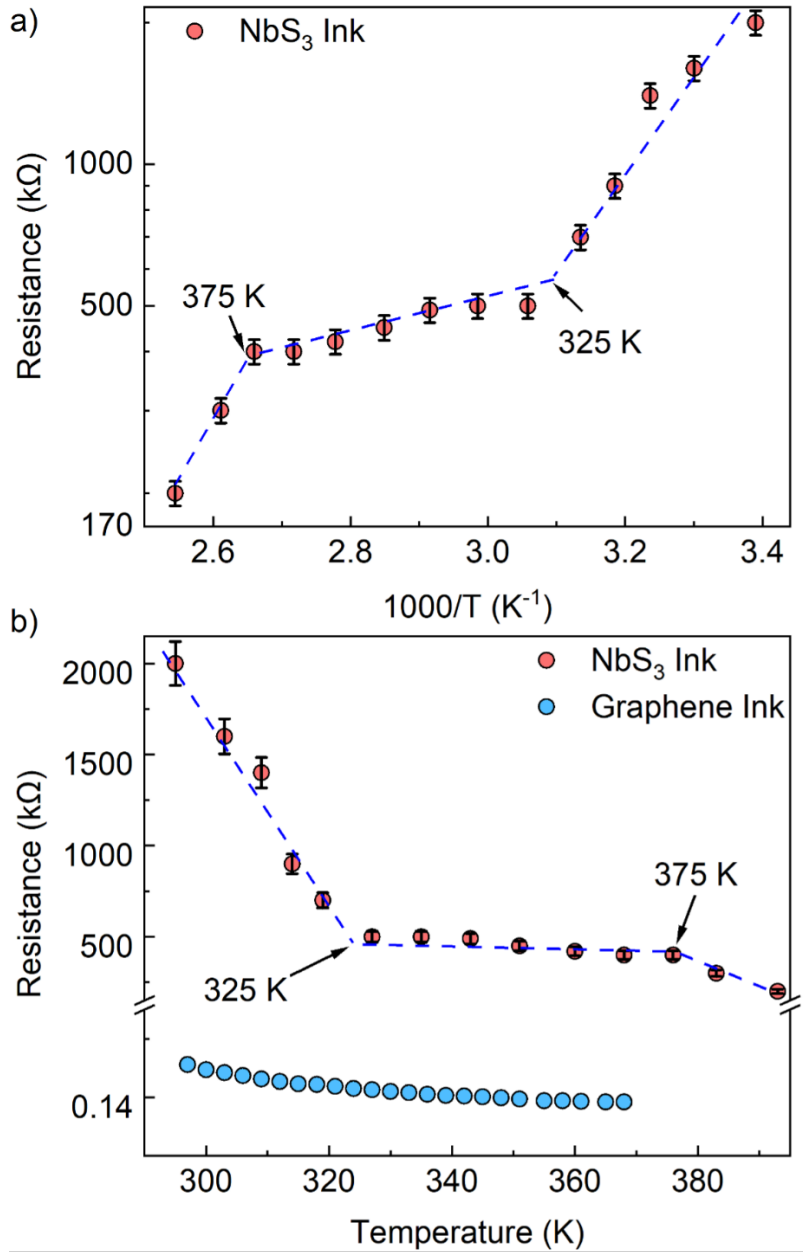

[**Figure 5**: (a) Resistance, *R,* of the printed device with NbS$_3$ ink. The data is plotted as a function of inverse temperature, 1000/*T,* in the semi-log scale. Note the onset of the transitions at T~325 K and T~375 K. (b) The *R vs. T* behavior of the reference graphene ink device for comparison with the NbS$_3$ ink device. The temperature-dependent resistance of the graphene ink shows a gradual monotonic drop with temperature rise without any noticeable deviations. In contrast, the NbS$_3$ ink exhibits an obvious change in resistive characteristics at T ~ 325 K and T ~ 375 K. This comparative approach confirms the changes in resistive behavior of the ink stem from the CDW phase transition of NbS3-II fillers present within the ink.]

## 5. CONCLUSIONS

In summary, we presented the development of functional "quantum inks" containing quasi-1D $NbS_3$ CDW fillers tailored for inkjet printing applications. The ink formulation involved liquid-phase exfoliation of bulk $NbS_3$ in DMF, followed by dispersing high-aspect ratio $NbS_3$ fillers in a mixture of ethylene glycol and IPA. Two-terminal devices were printed using the formulated $NbS_3$ ink, and graphene ink as a reference, on $Si/SiO_2$ substrates. Electrical resistance measurements of $NbS_3$ devices exhibited notable changes at T~330 K and T~370 K. Such changes in resistive behavior were not observed for the reference printed devices with graphene ink. The observed anomalies attributed to the CDW transitions of individual $NbS_3$ fillers of type-II present in the ink. Our findings validate that $NbS_3$-II maintains its intrinsic CDW properties despite undergoing rigorous chemical exfoliation processes. These outcomes are significant for advancing functional inks utilizing quasi-1D fillers with CDW properties, promising a wide array of applications in printed electronics.

## Acknowledgments

The work at UCLA was supported, in part, *via* the Vannevar Bush Faculty Fellowship (VBFF) to A.A.B. under the Office of Naval Research (ONR) contract N00014-21-1-2947 on One-Dimensional Quantum Materials. A.A.B. also acknowledges the support from the National Science Foundation (NSF) program Designing Materials to Revolutionize and Engineer our Future (DMREF) *via* a project DMR-1921958 entitled Collaborative Research: Data Driven Discovery of Synthesis Pathways and Distinguishing Electronic Phenomena of 1D van der Waals Bonded Solids.

## Author Contributions

A.A.B. and F.K. conceived the idea, coordinated the project, contributed to experimental data analysis, and led the manuscript preparations. T.G. developed the ink, printed devices, measured the electrical properties, and contributed to the data analyses; M.T. fabricated nanodevices, conducted SEM characterization, $I$-$V$ measurements, and contributed to data analyses; N.S. synthesized bulk crystals by CVT; S.G. contributed to the experimental data analyses; T.T.S supervised the material growth and contributed to data analysis. All authors participated in the manuscript preparation.

## The Data Availability Statement

The data that support the findings of this study are available from the corresponding author upon reasonable request.